\documentclass[preprint,aps,prl,showpacs]{revtex4}
\usepackage{graphicx}

\def\ub#1{{\bf #1}}

\begin{document}

\title{Superstructures at low spin - high spin transitions}

\author{D.~I.~Khomskii}

\affiliation{ II.Physikalisches Institut, Universit\"at zu K\"oln,
 Z\"ulpicher Str.77, 50937 K\"oln, Germany,
 and
 Laboratory of Solid State Physics, Groningen University,
 9747 AG Groningen, The~Netherlands}

 \author{U.~L\"ow}

\affiliation{Theoretische Physik, Universit\"at zu K\"oln,Z\"ulpicher
 Str.77, 50937 K\"oln, Germany}


\begin{abstract}
In many transition metal compounds, in particular those containing
Fe$^{2+}$ and Co$^{3+}$, there occur spin-state transitions between
low-spin and high-spin (or intermediate-spin) states. We show that
typical interactions between similar spin-state ions are short-range
repulsion, and long-range interaction which can have different sign
depending on the elastic anisotropy of the lattice and on the
direction between respective ions. Due to such character of effective
interactions at the spin-state transitions there may occur different
superstructures --- ordered arrangement of different spin-states,
which in particular may have the form of stripes. The properties of
the system TlSr$_2$CoO$_5$ for which such a superstructure was
recently observed experimentally, are discussed from this point of
view.
\end{abstract}

\maketitle

\pagebreak

There exist many transition metal compounds in which, due to a
competition between crystal field splitting and intra-atomic Hund's
rule exchange, different spin states lie close in energy. Such is
often the situation for Fe$^{2+}$ and Co$^{3+}$. For these ions the
low-spin (LS) state (occupation of d-levels t$_{2g}^6$e$_g^0$, S=0),
intermediate-spin (IS) state (t$_{2g}^5$e$_g^1$, S=1) and high-spin
(HS) state (t$_{2g}^4$e$_g^2$, S=2) can be stabilised at different
conditions. In some cases there occur spin-state transitions caused
e.g. by change of pressure, temperature or composition. There are
many examples of such transitions among Fe$^{2+}$-compounds
\cite{slichter,koenig}, in particular with organic ligands. Among
systems with Co$^{3+}$, the best known (and still controversial) is
the case of LaCoO$_3$. For a long time it was assumed that there
occurs in it the LS -- HS transition with increasing temperature
\cite{racah}, but more recent calculations \cite{korotin} and
experiments \cite{IS} point rather to the LS -- IS transition.

Usually spin-state transitions occur between homogeneous states
having predominantly one particular spin state; often they are of
first order. However as we shall show below, a quite
different and interesting situation may occur: the interaction
between ions with different spin state may be such that at the
spin-state transitions an intermediate state may be stabilised,
in which there appears {\sl a spin-state superstructure} - an ordered
arrangement of different spin states of the same ionic species,
e.g. an alternation of HS and LS states. Recently the first
example of such superstructure was observed in a layered
cobaltite  TlSr$_2$CoO$_5$ \cite{doumerc} and its presence was
suggested for NdBaCo$_2$O$_{5.5}$ \cite{Fauth}.

In the present paper we consider the
form of interactions determining the character of spin-state
transitions, suggest a general mechanism which can lead to
spin-state superstructures, and qualitatively explain the
superstructure observed in TlSr$_2$CoO$_5$.

Conventional description of spin-state transitions
\cite{sugano,spiering} relies on interactions induced by lattice
distortions. Ions with different spin states have different ionic
radii, e.g. Co$^{3+}$ in the LS state in VI-fold coordination has
the ionic radius 0.545 \AA, and in the HS state --- 0.61 \AA.
Consequently, if we change the spin state of an ion, e.g.
by transforming one of the LS Co's into a HS state, it amounts to
introducing an``impurity'' with larger size into a matrix (the so
called ``sphere in the hole'' model \cite{sugano}). This
introduces strain in a crystal, and coupling to this strain
provides the mechanism of interaction between impurities. In a
model of isotropic elastic continuum generally used in treating LS
-- HS transitions \cite{sugano,spiering}, the only interaction
remaining is that through the ``image forces'' due to a
stress-free surface of the sample \cite{eshelby}, which is
attractive. With such an attractive interaction one indeed obtains
homogeneous structures, transitions between which may be of first
order.

In real crystals, however, the elastic interactions are more
complicated. Thus e.g. in weakly anisotropic cubic crystals there
appears an interaction between impurities which decays rather
slowly, as $\frac{1}{R^3}$, and, most important, which has a
different sign in different directions \cite{eshelby,whystripes}:
\begin{equation}
 V(\vec r, \vec r \, ')=-Cd\, Q_1 Q_2 \frac{\Gamma(\vec n)}{|\vec
r-\vec
 r\,'|^3}
 \label{eq1}
 \end{equation}
 where $C$ is a constant of the order of unity, $\vec r-\vec r\,'=|\vec
r-\vec r\,'|\cdot\vec n$, \ $Q_1$ and $Q_2$
 are the ``strengths'' of impurities ($Q_i\sim(v_i-v_0)$ where $v_i$
 is the volume of the impurity and $v_0$ is the corresponding volume
 of the matrix), and
 the angular
 dependence of the interaction~(\ref{eq1}) is determined by a
 function of the direction cosines of the vector
 $\vec R=\vec r_1-\vec r_2$:
 \begin{equation}
 \Gamma(\vec n)=n_x^4+n_y^4+n_z^4-{\textstyle\frac35}.
 \label{eq2}
 \end{equation}
Here the elastic anisotropy is characterised by the parameter
$d = c_{11} - c_{12} - 2c_{44}$, where $c_{ij}$ are the corresponding
elastic moduli. We see that e.g. for $d<0$ the
interaction (\ref{eq1}) is positive (repulsive) along the 
[100], [010] and [001]
directions and attractive along face and body diagonals [110] and
[111], and vice versa for $d>0$. Thus we see that, in contrast to
usual assumptions, the interaction between similar states (e.g.
between HS ions in a LS matrix) may be repulsive, at least in certain
directions.

One can also show that the nearest neighbour (nn) interaction via
short-wavelengths or optical phonons is typically also repulsive.
Thus e.g. in a perovskite lattice, when one puts a large HS ion at a
certain site, it pushes apart the surrounding oxygens,
so that it is more favourable to have small LS ions at the nn
 sites.
Consequently, the effective interaction between similar ions at the
nn sites is repulsive (or equivalently, it is attractive between LS
and HS states). As is clear from these arguments,
one can expect in this case that not only homogeneous states (LS
or HS), but also certain states with ordered LS and HS ions can
appear at certain conditions.

A convenient way to describe spin-state transitions and eventual
superstructures is to map this situation to an effective lattice
gas, or Ising model. Let us consider the situation in which each
ion can be in two states, e.g. LS and HS (it may be also LS and IS
\cite{korotin,IS} or, in the case of TlSr$_2$CoO$_5$, IS and HS
states \cite{doumerc}; for simplicity we speak below about LS and
HS). We introduce pseudospin operators $\sigma_i = \pm 1$, so that
$\sigma_i = -1$ corresponds to a LS state, and $\sigma_i = 1$ 
to a HS state at site $i$. According to our general discussion
we can model the situation by the effective Ising-type interaction
$\sum J_{ij} \sigma_i \sigma_j$, containing in general
short-range repulsion, and a long-range part which decays
as $\frac{1}{R^3}$ and which may have different signs in
different directions; this longer-range interactions depend on
the details of the crystal structure, elastic anisotropy etc., see
Eq.(\ref{eq1}). The relative energies of different spin-states,
e.g. HS vs LS, will be described in this language by the
effective "magnetic field" --- the term $h\sigma_i$.

 It is known that the Ising model with long-range interactions can
give rise to a variety of different ordered structures (see the
well-known ANNNI model \cite{selke}  or the
treatment of the 2d Ising model with the ``Coulomb'' interaction in
\cite{loew}). In this paper we consider a simplified model,
which nevertheless contains the essential physics,
keeping only a small number of pair interactions. In contrast to
the treatment of \cite{loew,whystripes}, where one has studied
systems with fixed concentration of particles (in our mapping ---
fixed ``magnetisation''), in our present problem the relative number
of different spins (LS and HS ions) is not fixed. Consequently we
should consider our system not for fixed density (``magnetisation''),
but for fixed chemical potential (``magnetic field''); the role of
the temperature can be also mapped onto a magnetic field \cite{sugano}.

We consider a 2d square lattice, modelling the situation in
TlSr$_2$CoO$_5$, which is the layered compound with
perovskite-like CoO$_2$-plane similar to CuO$_2$-plane in
high-T$_c$ cuprates. The Hamiltonian of our model is

 \begin{equation}
 H = J_1\!\!\sum_{\langle ij\rangle=nn}\!\!\sigma_i\sigma_j
  + J_2\!\!\sum_{\langle ij\rangle=nnn}\!\!\sigma_i\sigma_j-\rm h
\sum_i \sigma_i
 \label{eq3}
 \end{equation}

Here the first term describes the nearest neighbour interaction,
which, as we argued above, is repulsive, $J_1>0$; the second term is
the next nearest neighbour interaction along $x$ and $y$-directions
([100] and [010]) which we also take to be repulsive.
The last term in Eq.(\ref{eq3}) describes the difference
of the on-site energies of LS and HS states, which can be changed
e.g. by pressure etc. (For Co$^{3+}$ the effective field $h$ will be
equal to $h=6J_H - 2 \Delta$, where $J_H$ is the Hund's rule exchange
and $\Delta$ is the crystal-field splitting between t$_{2g}$ and
$e_g$-levels).

It is straightforward to see that for only nn repulsion ($J_1>0$,
$J_2 = 0$) three states can be realized at $T=0$ for different values
of $h$: the state with all $\sigma_i=-1$ (LS-states) and with the
energy (per site) $E_{\text{LS}}=\frac{J_1 z}{2}+h$ where $z$ is the number of
nearest neighbours (in our case $z=4$); the state with alternating
spins $+1$,$-1$,$+1$, \dots , forming a two-sublattice
``antiferromagnetic'' structure in a bipartite lattice (we consider
below only such a case), $E_{\text{LS/HS}} = -\frac{J_1 z}{2}$, and the
homogeneous HS state with all $\sigma_i=1$ and
$E_{\text{HS}}=\frac{J_1 z}{2}-h$. Consequently, we would have jump-like phase
transitions between these states with increasing $h$, from the LS
state at $h<-J_1 z$ to an ordered array LS/HS/LS/HS~\dots~for
$-J_1 z<h<J_1 z$, and to a HS state for $h>J_1 z$. Thus we see that this model
quite naturally leads to the formation of a state with a
superstructure of LS and HS states, similar to the one observed in
\cite{doumerc}.

One important difference of this simple case and the experimental
situation observed in \cite{doumerc} is that the ratio of LS to HS
states (IS to HS in the real case of TlSr$_2$CoO$_5$) is not 1:1
as above, but 1:2, IS states forming diagonal stripes in the 2d
square lattice. But just such a state appears when one takes into
account the second term in the Hamiltonian Eq.(\ref{eq3}).

In order to determine the ground state phase diagram   
of  the model (\ref{eq3}) we employed
a Metropolis Monte Carlo algorithm combined with a single spin flip 
dynamics to cool the states to zero temperature \cite{loew}. 
We further compared the energies of all possible 
periodic states with unit cells up to size $5\times 5$.
In this way we found four regions with 
different ordered ground states:
a ferromagnetic and an antiferromagnetic phase,
a $2\times2$ checkerboard structure
which is degenerate with  $(2,2)$-stripes, and a $(2,1)$ stripe phase.
The resulting phase diagram for $h>0$ is shown in Fig.1
(as is clear from the form of the Hamiltonian (\ref{eq3}), 
the phase diagram for $h<0$ can be obtained by changing $\sigma$ to $-\sigma$, i.e. by reflecting
the phase diagram in Fig.1 relative to the x-axis and changing  
$+ \leftrightarrow\ -$).

Since all the ground states are simple diagonal 
stripes (except for the checkerboard phase which is however 
degenerate with the $(2,2)$ stripe phase)
we mapped the model onto a one-dimensional
effective model, which is simpler 
to analyse and which served us for a detailed 
analysis (see appendix). This study confirms the conclusions presented
above and reproduces the phase diagram of Fig.1.

To summarize, away from  the boundaries all systematic checks revealed 
no phases but the ones shown in Fig.1.
Note however, that 
on the boundary lines interesting degeneracies occur. 
Thus e.g. on the line $h=4J_1+4J_2$  separating the ferromagnetic 
and the (2,1) phase  all diagonal 
stripes (n,1) consisting of n spins  $\sigma =+$ 
and one spin $\sigma=-1$ are degenerate.
We stress however that these degeneracies are of no phenomenological 
relevance for the superstructures studied here, since 
they are strictly confined to the "one-dimensional" phase boundaries.
 
The phase diagram obtained above resembles that of the well-known
ANNNI model, cf. Fig.1 in Ref.\cite{uimin}, with the difference
that because of the ferromagnetic coupling between nn spins in
one direction in the ANNNI model, $(2,1)$ stripes in the latter are
vertical and not diagonal, as in our case (see the appendix for details).
Also, our result agrees with Ref.\cite{Landau}
where the model Eq.(\ref{eq3}) is discussed for $h=0$.

From the results presented above we see that in a large part of the
phase diagram we obtain the phase with (2,1) diagonal stripes; this phase 
exactly corresponds to the spin-state superstructure observed 
experimentally in TlSr$_2$CoO$_5$ \cite{doumerc}.

We also considered the generalized model, including the
interaction between the sites along diagonals in the plaquette,
which according to Eqs.(\ref{eq1},\ref{eq2}) should be attractive
(ferromagnetic): the corresponding model in the spin language is

\noindent
\begin{eqnarray}
H = J_1 \sum_{<ij>=nn}\sigma_i \sigma_j 
+ J_2 \sum_{<ij>=nnn}\sigma_i \sigma_j 
+  J_d \sum_{<ij>=diag}\sigma_i \sigma_j 
-h \sum_i \sigma_i.
\label{eq6}
\end{eqnarray}
\noindent

where $J_1,J_2>0$, $J_d<0$.

We found no new ground state phases emerging for this case, but the effect of
the diagonal coupling $J_d$ is merely to shift the phase boundaries
in Fig.1 by $J_d/2$ to the right (see appendix).

For a longer range interaction, e.g. the one of the type (\ref{eq1}),
certain other states may appear, depending on the ratio of the
constants of the Hamiltonian. The description of these states is a
formidable problem even in 1d-case, especially at finite temperatures
\cite{selke} (see also the results of numerical calculations in
\cite{loew}). We will not discuss all the details of this problem
here; our main aim is just to demonstrate the possibility of the
appearance of states with spin-state superstructures at
corresponding transitions. We see that such superstructures indeed
occur quite naturally if we take into account realistic interactions
of HS and LS states via lattice deformation, including elastic
anisotropy and interaction with optical phonons. In particular, the
superstructure observed experimentally in TlSr$_2$CoO$_5$ is stable
for a certain range of parameters if we include nn and nnn interaction
along [100] and [010]-directions.

Concerning the details of the properties of TlSr$_2$CoO$_5$ and
of the theoretical description thereof, a few extra points should be
mentioned. We discussed above only the origin and type of
ordering in the ground state, assuming localised electrons. In
reality TlSr$_2$CoO$_5$ undergoes the first order insulator -- metal
transition with increasing temperature at about 300 K. We do not
discuss this transition in this paper, but one can argue that the
energy gap in the low-temperature phase may be connected with the
occurrence of the spin state superstructure. From this point of
view one probably may consider this state as a result of the
formation of the spin-state density wave (SSDW), starting from
the high-temperature metallic phase. (The treatment presented
above would correspond to a strong-coupling limit of this
picture.)

Another point worth mentioning is the suggestion \cite{foerster}
that an orbital degeneracy may play a role in the properties of
TlSr$_2$CoO$_5$. In general, indeed, there may exist
$e_g$-degeneracy in the IS state and $t_{2g}$-degeneracy in both
the IS and HS states of Co$^{3+}$. However from the experimental
data \cite{doumerc} it follows that this orbital degeneracy is
predominantly lifted due to a rather strong tetragonal elongation
of the CoO$_6$-octahedra which exists already in the
high-temperature phase and which most probably is connected with
the layered structure of this compound. Indeed, the IS state for
this distortion is nondegenerate, and only in the HS state there
remains a double degeneracy in the $t_{2g}$-sublevels. But the latter
in Co compounds is usually lifted by the spin-orbit coupling,
which presumably would be the case here as well (Jahn-Teller
effect by itself would stabilize for the HS state not an elongation,
but a compression of CoO$_6$-octahedra). This question, however,
deserves further study.

In conclusion, we considered in this paper the properties of
systems undergoing spin-state transitions, like the ones often
observed in materials containing Fe$^{2+}$ and Co$^{3+}$. We argued
that the effective interaction governing the behavior of such
systems -- the interaction via lattice distortions -- is more
complicated than the usually assumed attraction between similar
spin-state ions: typically the nearest neighbour interaction is
repulsive, and more distant interactions may be of either sign,
depending on the elastic anisotropy of the crystal and on the
direction between respective ions. As a result of such form of
interaction, different superstructures, consisting of ordered
distributions of different spin-states, can naturally occur in this
case. We discussed from this point of view the spin-state
superstructure observed in the low-temperature phase of TlSr$_2$CoO$_5$
\cite{doumerc}, and have shown that this superstructure can be
explained by our model for a certain range of parameters. It would be
interesting to look for similar superstructures in other systems
with comparable energies of different spin-states and with spin-state
transitions.

\section{appendix}

To gain further insight and to corroborate the picture 
obtained by Monte Carlo calculations and by explicitly 
comparing the energies of different states,
we took advantage of the in principal one-dimensional structure of
the relevant configurations (i.e. diagonal stripes, see Fig.1) and
mapped the original Hamiltonian (\ref{eq6})
onto a one-dimensional Ising model in a magnetic field
with the Hamiltonian:

\begin{eqnarray}
H_{\text{1dim}}= \sum_i \{ 2J_1 s_i s_{i+1} + (2 J_2 + J_d) s_i s_{i+2} 
-h s_i +  J_d \} .
\label{one}
\end{eqnarray}
\noindent

Here $s_i=\pm 1$  represents the normalized sums of the spins 
on the diagonals and the one-dimensional interaction is orthogonal 
to the stripes. Note that this Hamiltonian contains all possible 
diagonal stripe configurations. Using this one-dimensional model we
checked systematically for possible diagonal stripe configurations
up to  unit cell of 18.

From Eq.\ref{one} one sees that --- once diagonal 
stripes are confirmed as ground states ---  
the role of $J_d$ is a mere modification of the 
next nearest neighbour coupling (plus an irrelevant constant)
and hence a shift of the phase boundaries by $J_d/2$.

Eq.\ref{one} is also useful to display the similarities to the 
two-dimensional ANNNI-model 

\begin{eqnarray}
H_{\text{ANNNI}} = - \sum_{x,y} ( \tilde J_1 S_{x,y} S_{x+1,y} 
+ \tilde J_2  S_{x,y} S_{x+2,y} 
+  \tilde J_0  S_{x,y} S_{x,y+1} 
+ H  S_{x,y} )
\label{ANNNI}
\end{eqnarray}

\noindent
with $\tilde J_1,\tilde J_2<0$  and $\tilde J_0>0$ 
as discussed in Ref.\cite{uimin}.
Due to the ferromagnetic interaction $\tilde J_0$ in $y$-direction,
the stripes are vertically oriented in the case of Eq.\ref{ANNNI}
and the corresponding one-dimensional Hamiltonian 
is

\begin{eqnarray}
\tilde H_{\text{1dim}}=- \sum_i \{ \tilde J_1 s_i s_{i+1} + \tilde J_2 s_i s_{i+2} 
+H s_i +  \tilde J_0 \}. 
\label{one_ANNNI}
\end{eqnarray}

$\tilde H_{\text{1dim}}$ Eq.(\ref{one_ANNNI}) differs 
from $H_{\text{1dim}}$ Eq. (\ref{one}) for $J_d=0$ 
by a factor 2 in the
couplings and by a constant. The phase diagrams of
the models can thus be mapped onto one another 
by a simple rescaling of the axis.

\subsection{Acknowledgments}

We are grateful to J.~P.~Doumerc, D.~F\"orster and R.~Hayn for
useful discussions. This work was supported by the 
Sonderforschungsbereich 608 of the Deutsche Forschungsgemeinschaft,
by the Netherlands Foundation for Fundamental Study of Matter 
(FOM) and by the
Netherlands Science Organisation (NWO).

\pagebreak

\pagebreak

\begin{figure}
\includegraphics[width=\linewidth]{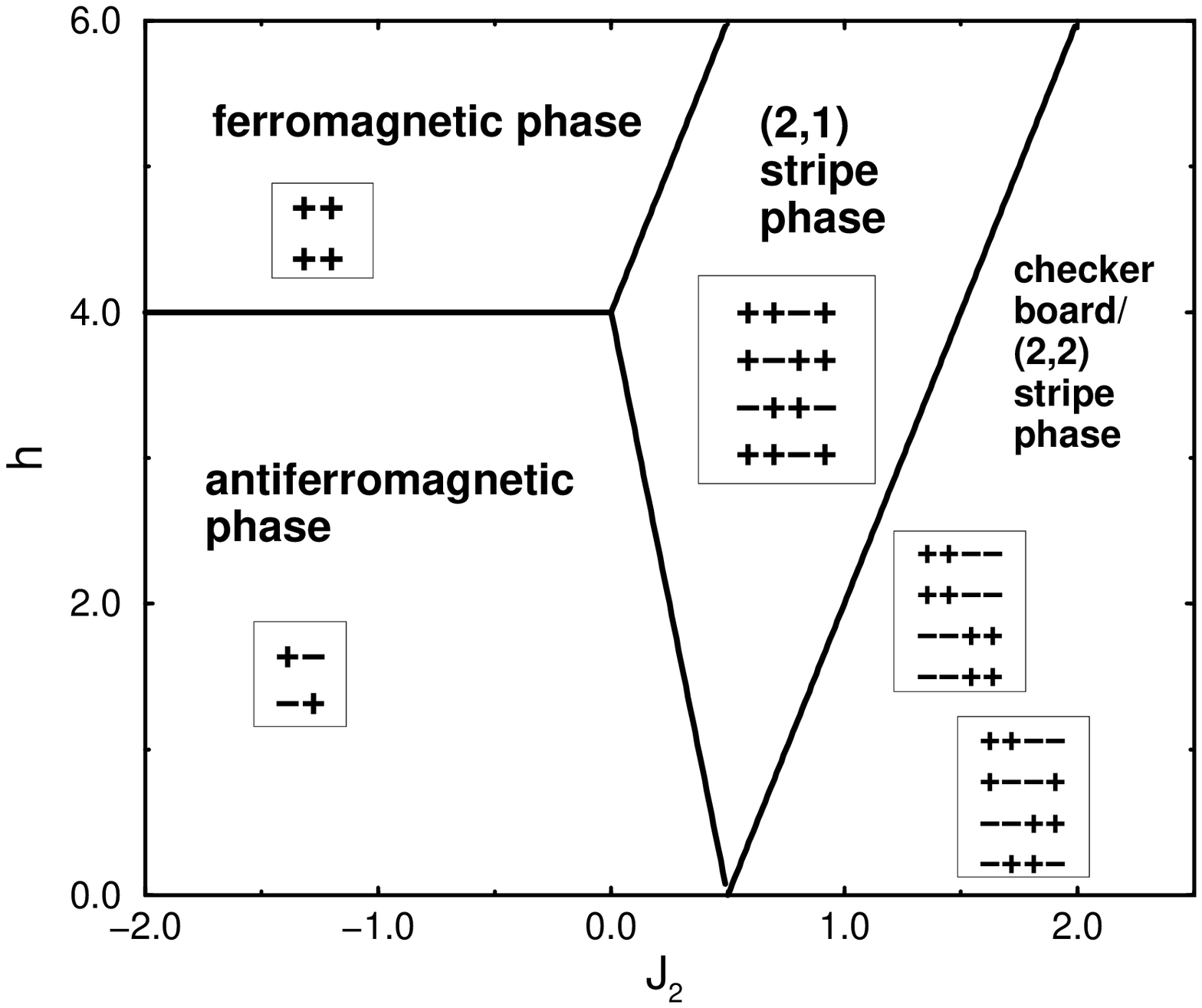}
\begin{center}
\caption{Phase diagram of the model (\ref{eq3}). The type of pseudospin 
ordering is shown for each possible phase. (The sign $+$ corresponds 
e.g.to a high-spin state, and  the sign $-$  to a low-spin or 
intermediate-spin state.) Note, that we fixed the
energy scale by putting $J_1=1$.}
\end{center}
\end{figure}

\end{document}